\documentclass[pre,twocolumn,preprintnumbers]{revtex4}
\usepackage{amssymb}
\usepackage{amsmath}
\usepackage{graphicx}
\usepackage{makecell}
\usepackage{multirow}
\usepackage{float}
\usepackage{lipsum}
\usepackage[caption = false]{subfig}
\begin{document}

\title{Dynamical and Topological Aspects of Consensus Formation in Complex Networks}

\author{A. Chacoma}
\author{M. N. Kuperman}
\author{G. Mato }

\affiliation{Centro At\'omico Bariloche (CNEA) and Instituto Balseiro}
\affiliation{Consejo Nacional de Investigaciones Cient\'{\i}ficas y T\'ecnicas, \\
(8400) Bariloche, R\'{\i}o Negro, Argentina.}

\begin{abstract} The present work analyses a particular scenario of consensus formation, where 
the individuals navigate across an underlying network defining the topology of the walks.  The 
consensus, associated to a given opinion  coded as a simple messages, is generated by interactions 
during the agent's walk  and manifest itself in  the collapse of the  various opinions into a single one. 
We analyze how the topology of the underlying networks  and the rules of interaction between the agents 
promote or inhibit the emergence of this consensus. We find that non-linear interaction rules are required to 
form consensus and that consensus is more easily achieved  in networks whose degree distribution is narrower.

\end{abstract}

\maketitle

\section{Introduction}

The role of complex networks as a  mathematical tool to formalize the underlying topology in 
several propagating phenomena of social extraction has been well established in the last years.
When looking for models of the spread of diseases or the propagation of  information,  
rumors and ideas, complex networks provide a plethora of alternative topologies that serve to mimic 
the complex weave of the interpersonal relationships \cite{bocca, kup1,kup2,kup0}. At a more 
abstract level, diffusion in general is one of the fundamental processes taking place in networks 
\cite{jack,lopi,garde}.  Diffusive propagation on a network generates the opportunity that agents 
get in touch and interchange ideas and information. Given the appropriate rules for interchange and 
network topology this could lead to the appearance of consensus among the opinions of the different 
agents.

Consensus formation is a widely studied phenomenon in social sciences. One of the main goals is to 
understand the 
emergence of  consensus in  a system involving a number of interacting agents \cite{wang-shang2015}. 
This problem has been studied analytically, for instance, in 
\cite{deGroot1974,Berger1981,GilardoniClayton1993}, where it is proposed that each agent alters its opinion according to some 
weighted average of the rest of system. It is found that if all the recurrent states of the Markov chain communicate with 
each other and are aperiodic, then a consensus is always reached. In most of the existing models the update of opinions
takes place via a linear mechanism \cite{sznajd-weron2000,holley1975,demarzo2003}. Moreover, they do not usually analyze the 
influence of the network topology. Although see, for instance, \cite{wang-shang2015}, where a complex network appears as a 
substrate of a set of agents with linear dynamics and \cite{fazeli2011}, where the topology of the network is altered by the 
interactions between the agents.
The structure of the complex networks has been extensively studied from the point of view of 
transfer of information. A lot of emphasis has been put on the influence of topology, studying for 
instance whether scale free networks are more  
efficient than regular ones \cite{sato}. In this context it has been shown, for instance, that it is very important 
whether the network consists of homogeneous nodes or it has a structure of routers and peripheral nodes. Another 
important aspect is the clustering of the nodes, because the presence or absence of loops  could affect information 
transfer \cite{sato}. For instance, in networks with modular structures, it has been shown that the velocity of the 
information propagation depends non linearly  on the the number of modules. A piece of information will propagate 
faster for networks having either a small number or a large number of modules \cite{huan}. 

A set of interacting agents on a network can give rise to a dynamically changing local environments where the process of 
interchange of opinions take place. In this paper we focus on three aspects: 1) How the local interaction rules control the 
convergence to consensus, in particular we analyze both linear and non-linear rules. 2) What is the influence of
dynamics of the agents in the network. We first consider the the propagation of the information by considering a 
na\"ive strategy for neighboring node selection. If the target node is not among the neighbors, a neighboring node is selected 
at random for the trajectory of the agent. Then we consider a preferential choice strategy, where the agents are more likely to 
move to more connected nodes. 3) What is the effect of the different parameters  of the network 
topology, such as clustering or assortativity. In the following sections we present the model in 
more details and a description of our results.

\section{Network Topologies}

Throughout this work we use several families of networks with  different topologies and 
algorithmic constructions, though always containing the same number of nodes and links.

\paragraph{\textit{Regular Small World networks:}} We consider first regular networks with a tuned degree of 
disorder, and consequently different degrees of clustering and mean path length. We recall that  
the clustering coefficient measures the tendency of 
the nodes to cluster together and can be locally characterized by the 
fraction of existent links between nodes in a given node neighborhood to the number of links that 
could possibly exist between them \cite{watt}.
The mean path length is the averaged path length between all te nodes in the network. By path 
length we define the minimum number of links neede to traverse to navigate from one node to he 
other.

Regular Small World networks are built using a modified algorithm based on the originally proposed 
in \cite{watt} to constrain the resulting networks to a subfamily with a delta shaped degree 
distribution. We call this family of networks the k-Small World Networks ($k$-SWN) \cite{kup3}, 
where 2$k$ indicates the degree of the nodes. 
The building procedure starts with an ordered regular network whose order is broken by 
exchanging the nodes attached to the ends of two  links  in a sequential way. Starting for example 
from an ordered ring network, each link is subject to the possibility of exchanging one of its 
adjacent nodes with another randomly chosen link with probability $p_d$. If the exchange is 
accepted, we  switch the partners in order to get two new pairs of coupled nodes. Double links are 
always avoided, thus if there is no way to avoid a double link with the present selection of nodes, 
a new choice is done. In this way all the nodes preserve their degree while the process of 
reconnection assures the introduction of a certain degree of disorder. 
It must be stressed that the dependence of the clustering coefficient $C$ and path 
length $L$ o the disorder parameter is qualitatively similar to the one  observed in Small World 
Networks \cite{watt}. In this way we can evaluate the effects of clustering and path length 
independently of the degree distribution.

\paragraph{\textit{Small World Networks:}} To include the possible effects of the degree 
distribution on the analyzed dynamics we consider then, the usual algorithm described in 
\cite{watt}, where only one link is rewired at a time, maintaining the attachment to one of the 
adjacent nodes and randomly connecting the other extreme. These networks not only posses different 
degrees of clustering and mean distance but also different binomial degree distributions, linked to 
the disorder parameter $p_d$, that measures the probability of changing the extremes of each link. 
The degree distribution of these networks changes 
from deltiform  to binomial at the moment of introducing the slightest disorder. The associated 
binomial distribution is given by \cite{barra}
$$P(h)=\sum_{n=0}^{\min[h,k]}C_k^n (1-p_d)^n p_d^{k-n} \frac{(kp_d)^{h-k-n}}{(h-k-n)\!}\exp(-kp_d),$$
where $k=K_m/2$ and $K_m$ is the mean degree of the network. We will refer to this networks as SWN.

\paragraph{\textit{Scale free networks:}} Finally, we are interested in studying networks with 
a degree distribution closely related to those frequently observed in real contexts. Thus we 
consider Scale Free Networks (SFN)  \cite{bara} with different degrees of assortativity, following 
the prescription presented in \cite{xulv} with a slight modification to obtain networks with 
tunable positive or negative assortativity. In order to get positive assortativity, the algorithm 
consists in departing from a random network with the desired degree distribution and  sequentially 
rearranging the nodes at the ends of a pair of randomly chosen links. At each step two links of the 
network are chosen at random. The corresponding four nodes are ordered with respect to their 
degrees and the original links are deleted. Then, with probability $p_a$,  a new link connects the 
two nodes with the smaller degrees and another connects the two nodes with the larger degrees. 
Otherwise, the links are randomly rewired. 
As double connections are forbidden, if the rewiring indicates connecting  already connected nodes, 
the change is  discarded and the original links are reestablished. If  instead of connecting the 
nodes as prescribed before  we connect the node with the highest degree 
with the node with the lowest degree, and the other two between them we obtain networks with 
negative assortativity. In both cases the parameter $p_a$ governs the final degree of assortativity.
The assortativity  $a$ can be measured as the Pearson correlation coefficient of degree between two 
neighboring nodes \cite{newa}. 
\begin{equation}
a=  \frac{\sum_{i} j_i k_i-M^{-1}[\sum_{i} \frac{1}{2}(j_i+k_i)]^2}{\sum_{i} 
\frac{1}{2}(j_i^2+k_i^2)-M^{-1}[\sum_{i} \frac{1}{2}(j_i+k_i)]^2},
\label{asortatividad}
\end{equation}
where the summation is over each of the $M$ links $i$ and,  $k_i$ and $j_i$ are the degrees of 
nodes at each of the ends of link $i$.
When this  coefficient is positive it indicates a higher occurrence of links between nodes of 
similar degree, while a negative value reveals connections between nodes of different degrees.

\section{Dynamics of consensus formation}

We have $N_p$ agents that travel from an initial node  of the network to a final one. These nodes 
are chosen
randomly with uniform distribution among the $N$ network nodes. Each agent has an opinion that is indexed by an integer
number between 1 and $N_c$. In the initial condition the opinion of each agent is chosen randomly with the constraint
that each opinion appears the same number of times. The number of agents is given by $N_p=10 N$, 
the number of nodes is $N=10000$ and the number of different opinions is $N_c=10$

At a given time there is some number of agents in a given node. If there are $N_i$ agents in node 
$i$, each one of them is 
assigned an index $j=1,\ldots,N_i$ according to the order of arrival, with earlier arrivals having 
lower indices. 
The node with the lowest index checks whether the node where is located is a neighbor of its destination node. 
If this is the case it moves in that direction and its movement stops. If not, it chooses one of its neighbors 
following one of two possible strategies. In one case, the node is randomly chosen among those conforming the 
neighborhood. The other strategy requires that each node has information about the degree of all its neighbors. 
The one with the higher degree is chosen, according to the preferential choice strategy \cite{kim} that a priori 
optimizes the possibility of reaching the target following a shorter path. This procedure is repeated in random order 
for all the nodes of the network.

This second strategy is motivated by the need to optimize the  ability to efficiently navigate and 
search in a network without a thorough knowledge  of its topological properties. As we are 
interested in finding the path to a target node from an initial one  using only local information, 
we appeal to the concept of random walk centrality \cite{noh} ($C_{rw}$), that characterizes those 
nodes which will be much frequently visited by initially uniformly distributed random walkers. 
Another interesting property is that when we consider the set of all possible random walks between 
two nodes, the one reaching  the node corresponding to a larger value of $C_{rw}$ is faster than 
the others. 
After finding that for a Barab\'asi-Albert scale-free network \cite{bara}, the degree of a node 
is directly related to its centrality the  maximum degree strategy was proposed in \cite{adam}. In 
this case, each node must have information on its neighbors’ degree so that a walking agent 
always moves to the neighboring node  having the highest degree. A relaxed version of this strategy 
was also proposed in \cite{kim}, where the authors also study the preferential choice strategy,  in 
which the node with the larger degree has the higher probability to be chosen. This strategy is 
equivalent to the former one when the probability is 1. 
As we will further show, these two strategies result in different results for consensus formation.

\subsection{Local interaction between agents}

We propose the following rules for the interaction between agents. The change of opinion will happen with 
probability $\rho$. In the present study we analyze two possibilities, namely: 

\begin{itemize}
 
\item{Linear:} An agent, instead of keeping its original opinion, takes a 
different one with probability $\rho$, randomly chosen among the opinions of the other agents in 
the same node.

\item{Quadratic:} An agent, instead of keeping its original opinion, takes 
two opinions randomly chosen among those present in the node, with probability $\rho$. If they are 
the same, then the agent will adopt it as the new opinion, otherwise, the original opinion 
is kept.

\end{itemize}

\subsection{Characterization of consensus formation}

These mechanisms for interaction lead to a global modification of the probability distribution of the opinions. 
This effect can be quantified using the Kullbak-Leibler distance \cite{kul}. Given two probability distributions 
$P$ and $Q$ it is defined by
\begin{equation}
 D_{KL} = \sum_k P(h) \log \frac{P(h)}{Q(h)}
\end{equation}
If the initial distribution is $Q(h)=1/N_c$ then the maximum value the Kullback-Leibler distance can 
take is when the distribution $P(h)$ is concentrated in only one point: $P(h)=\delta_{h,h_0}$, in 
which case $D_{KL}=\log(N_c)$. In other words consensus is maximized when $D_{KL}$ increases. Let us 
remark that the probability distribution is measured over the agents that have reached their 
target destinations. We report the normalized K-L distance, i.e., $D_{KL}/\log(N_c)$.

\subsection{Probability distribution of contents in the linear case}

We now prove that in the linear case the interactions between agents cannot alter the distribution 
probability of opinions. Let us denote by $P(k)$ the probability of having an agent with opinion 
$k$. The probability that this opinion is altered from $k$ to $l$ is $P(k\rightarrow l)= \rho 
P(l)$.
Similarly $P(l\rightarrow k)= \rho P(k)$. The probability of having an agent with  opinion $k$ (resp. $l$) 
in the first position of the node is $P(k)$ (resp. $P(l)$). Therefore the average number of agents that change 
their content from $k$ to $l$ is in average $N_p P(k) P(k\rightarrow l)= N_p \rho P(k) P(l)$ that is exactly
the number of opinions that change from $l$ to $k$. This implies that the Kullback-Leibler distance 
should not be affected by the interaction and consensus can never be achieved in the limit of large size. Any
change in the distribution is due to finite size fluctuations.

\begin{figure}[ht]
\begin{center}
\includegraphics[width=\columnwidth]{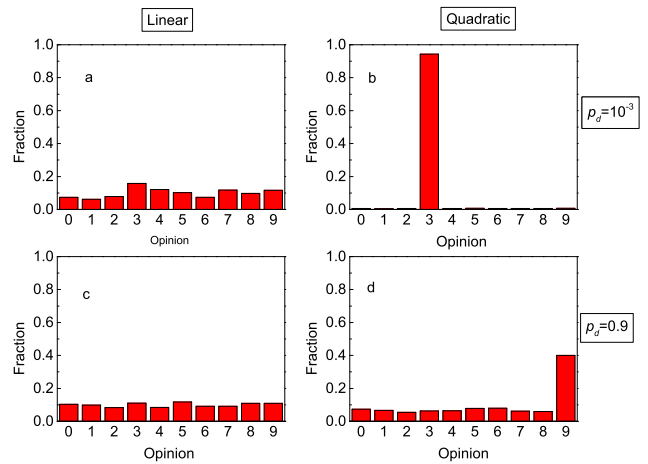}
\end{center}
\caption{Final distribution of opinions for two different dynamics and two disorder parameters of 
$k$-SWN's: a) Linear, $p_d=10^{-3}$, b) Quadratic, $p_d=10^{-3}$, c) Linear, $p_d=0.9$, d) 
Quadratic, $p_d=0.9$ }
\label{dischan}
\end{figure}

\begin{figure*}[!ht]
  \centering
  \subfloat[]{\includegraphics[width=0.5\textwidth]{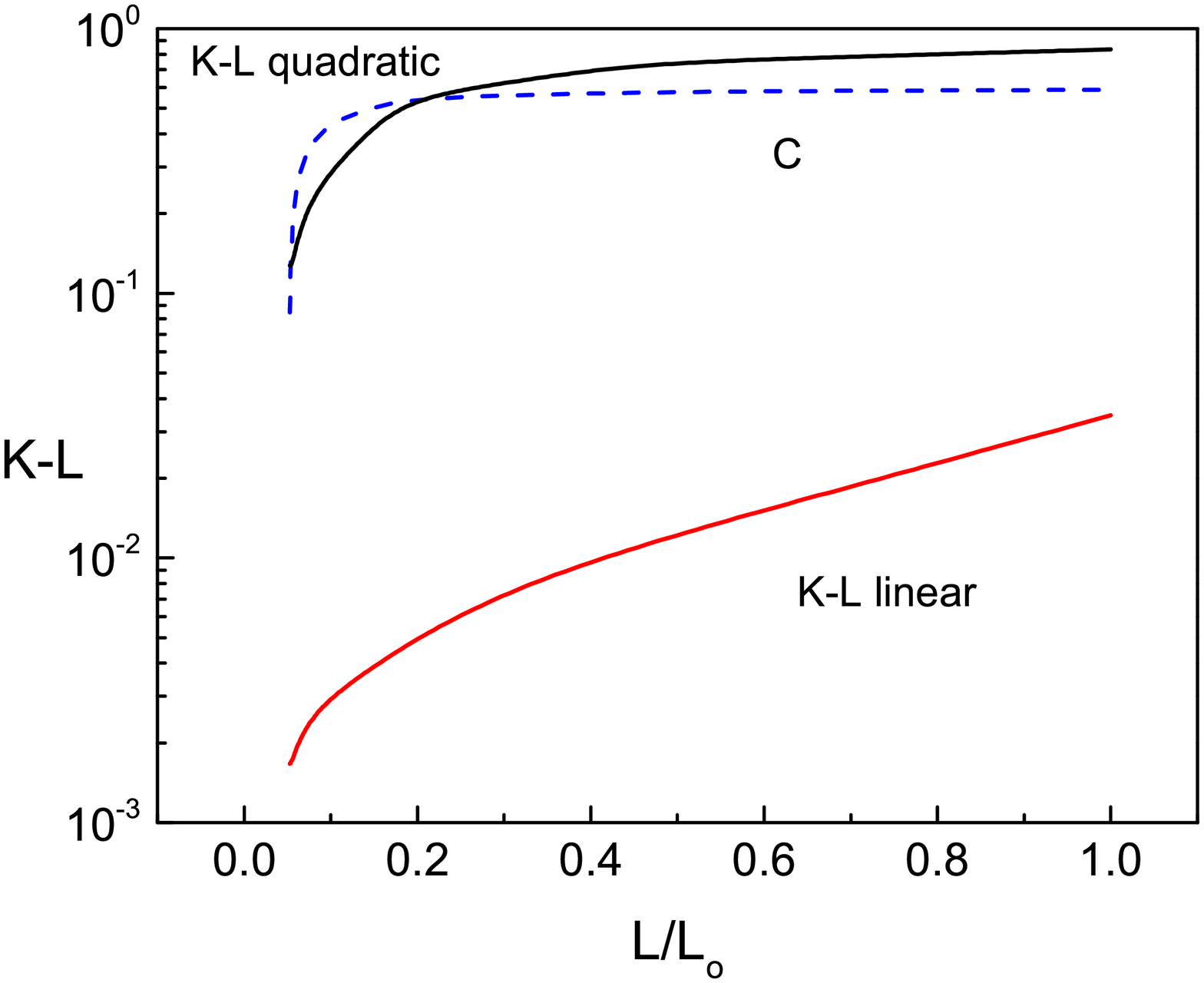}\label{dkl:f1}}
  \hfill
  \subfloat[]{\includegraphics[width=0.5\textwidth]{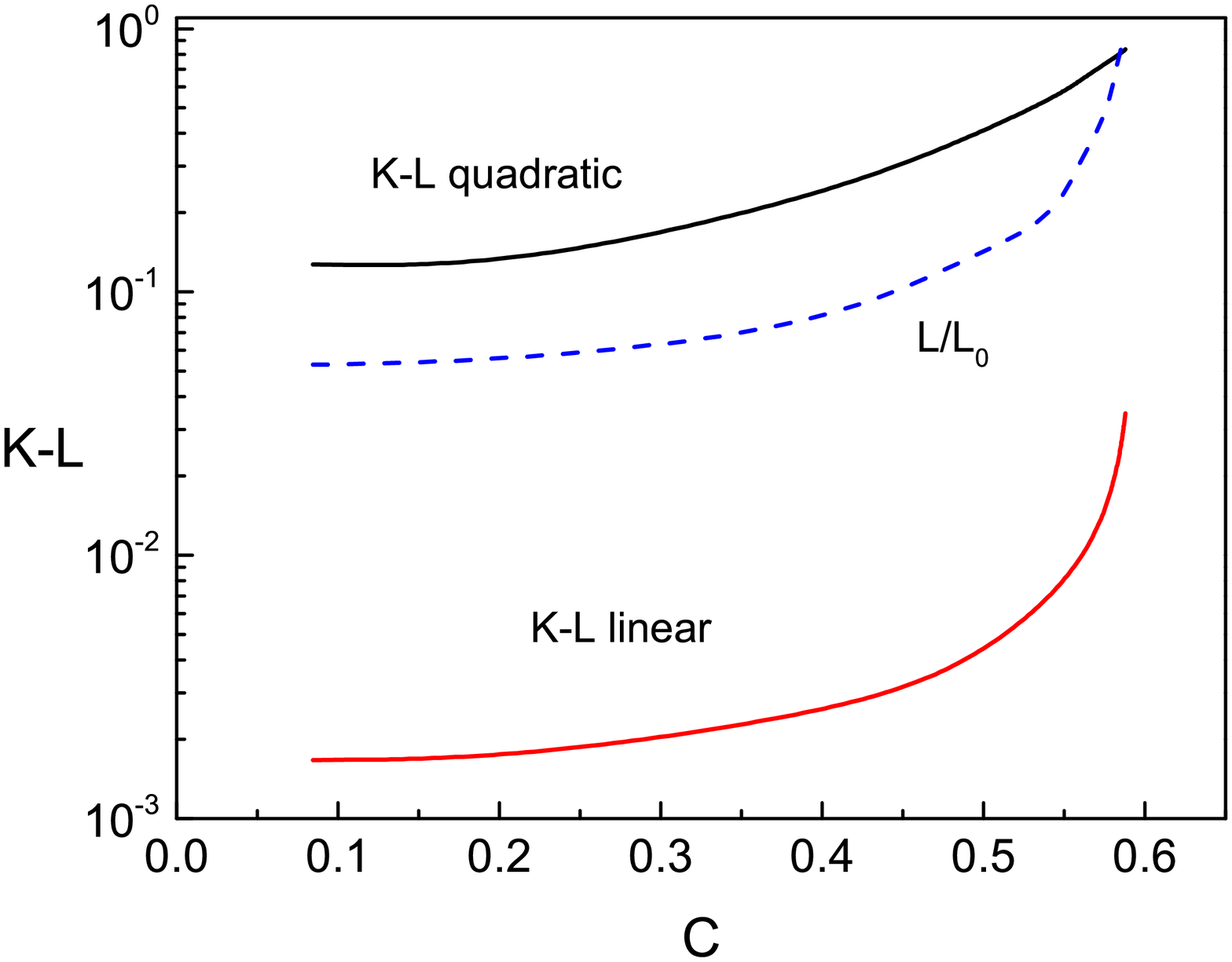}\label{dkl:f2}}
  \caption{Normalized K-L distance, $D_{KL}/\log(N_c)$, for different values of $p_d$ on $k$-SWNs and with linear and quadratic 
dynamics. a) as a function of $L/L_o$, with $L_o$ the mean path length of the ordered network. The 
dashed line is the value of $C$; b) as a function of $C$. The dashed line is the value of $L/L_o$. }
\end{figure*}

\section{Numerical results}

Although different network topologies were used throughout our simulations, all of them consisted 
in $N=10000$ nodes and a mean degree $K_m=6$. The results were averaged over 500 realizations each. 
We will discuss each topology separately and contrast the obtained results whenever the comparison 
is relevant. We took $\rho=0.02$, that is within the range where a high interaction rate does 
not screen the dynamics. At higher values it can be observed a too fast convergence to consensus. 
On the other hand it is
large enough to allow for enough interactions generate an effect in a reasonable time. The 
following results correspond to a navigation process that does not take into account the  degree of 
the neighboring nodes at the moment of choosing the next steps. The differences with respect to 
another  choice will be discussed later for SF networks.

\subsection{Final state of the dynamics}
\subsubsection{$k$-SWN}
As mentioned before, we took $N=10^5$, $K_m=K=2k=6$, and a disorder parameter $p_d$ 
ranging between 0 and 1. In order to characterize the formation of consensus we 
started by analyzing the change in the initially uniform distribution of messages.
As expected, when the underlying dynamics is linear, there is a slight departure from 
the initial distribution due to unavoidable fluctuations. On the contrary, when nonlinear dynamics 
are considered the situation changes dramatically. This can be observed in Fig. \ref{dischan}, 
where we show the final distribution of opinions for the linear and quadratic dynamics, for two 
limiting cases, $p_d=10^{-3}$ and $p_d=.9$

The departure of the  distribution of opinions from the initial one can be measured 
through the K-L distance. The difference between linear and quadratic dynamics are apparent 
throughout the whole range of $p_d$. In both cases, as the network disorder increases, the K-L 
distance decreases. In the linear case there is an increase but the distance is always at least 
one order of magnitude smaller than in the quadratic case. For the 
quadratic dynamics, it is possible to find an obvious reason in that as the mean path between nodes 
is smaller for disordered networks, a given agent takes less time to reach its target. This effect 
can be observed in Fig. \ref{dkl:f1}, where the values of the K-L distances are plotted against the 
average path length $L$. However, $L$ is not the only quantity that changes as a function of $p_d$. 
The clustering coefficient $C$ also decreases as the disorder grows and can be affecting the 
behavior of the K-L distance. The plot in 
Fig \ref{dkl:f2} shows the change in the K-L distance as a function of $C$. In both cases we 
observe two different regimes with slow and fast rates of change as a function of $L$ 
and $C$ respectively. We note that the fast change of the K-L distance in Fig. \ref{dkl:f1} 
corresponds to a range in which $C$ also changes rapidly as a function of $L$. The analogous 
situation is observed in Fig. \ref{dkl:f2}. We conclude that not only the change in path length 
contributes to the difference in the K-L distance but also the change in clustering. So far we have 
isolated the possible effect of a change in the degree distribution. In the next section this will 
be also taken into acount.

\begin{figure*}[!ht]
  \centering
  \subfloat[]{\includegraphics[width=0.5\textwidth]{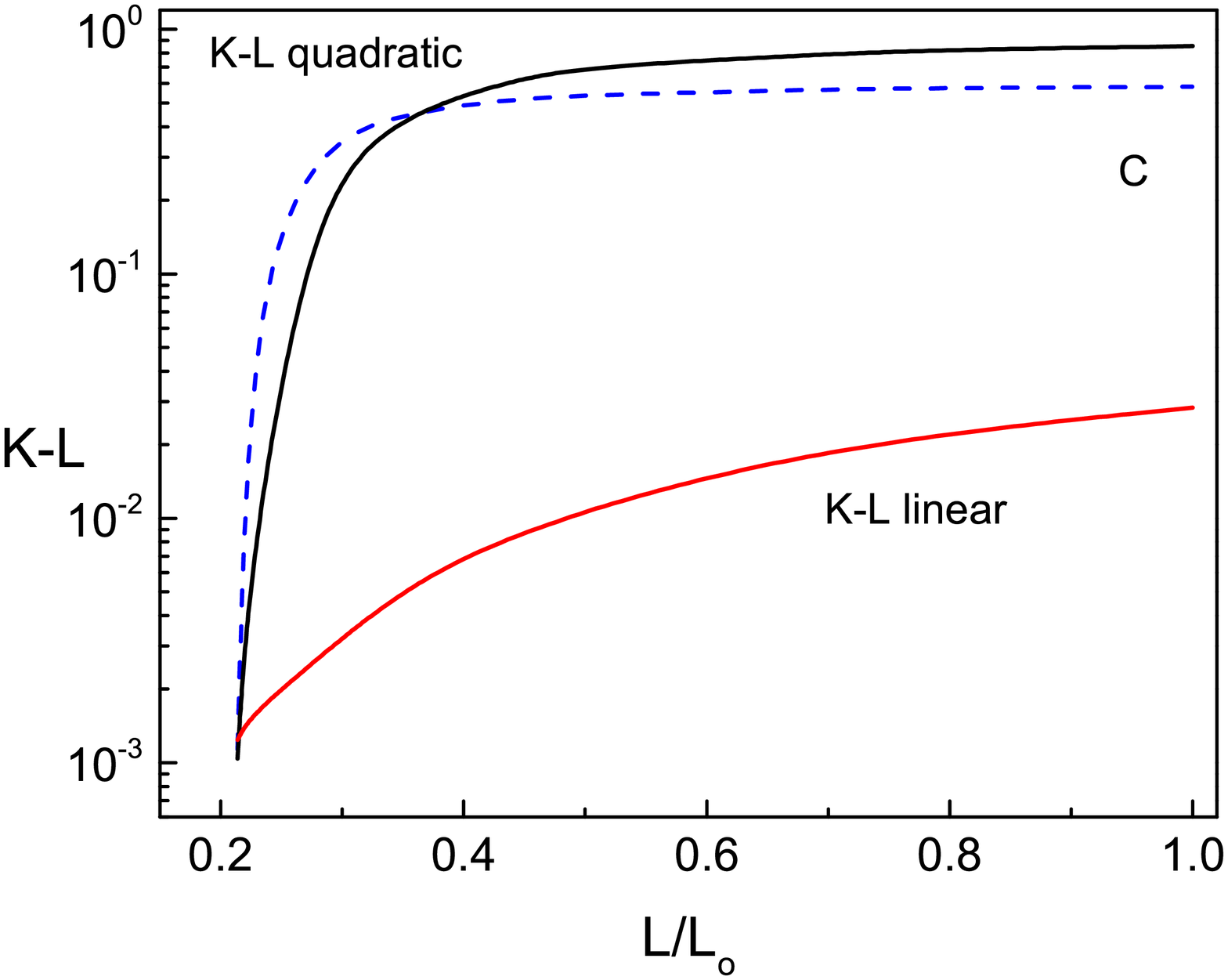}\label{dkl2:f1}}
  \hfill
  \subfloat[]{\includegraphics[width=0.5\textwidth]{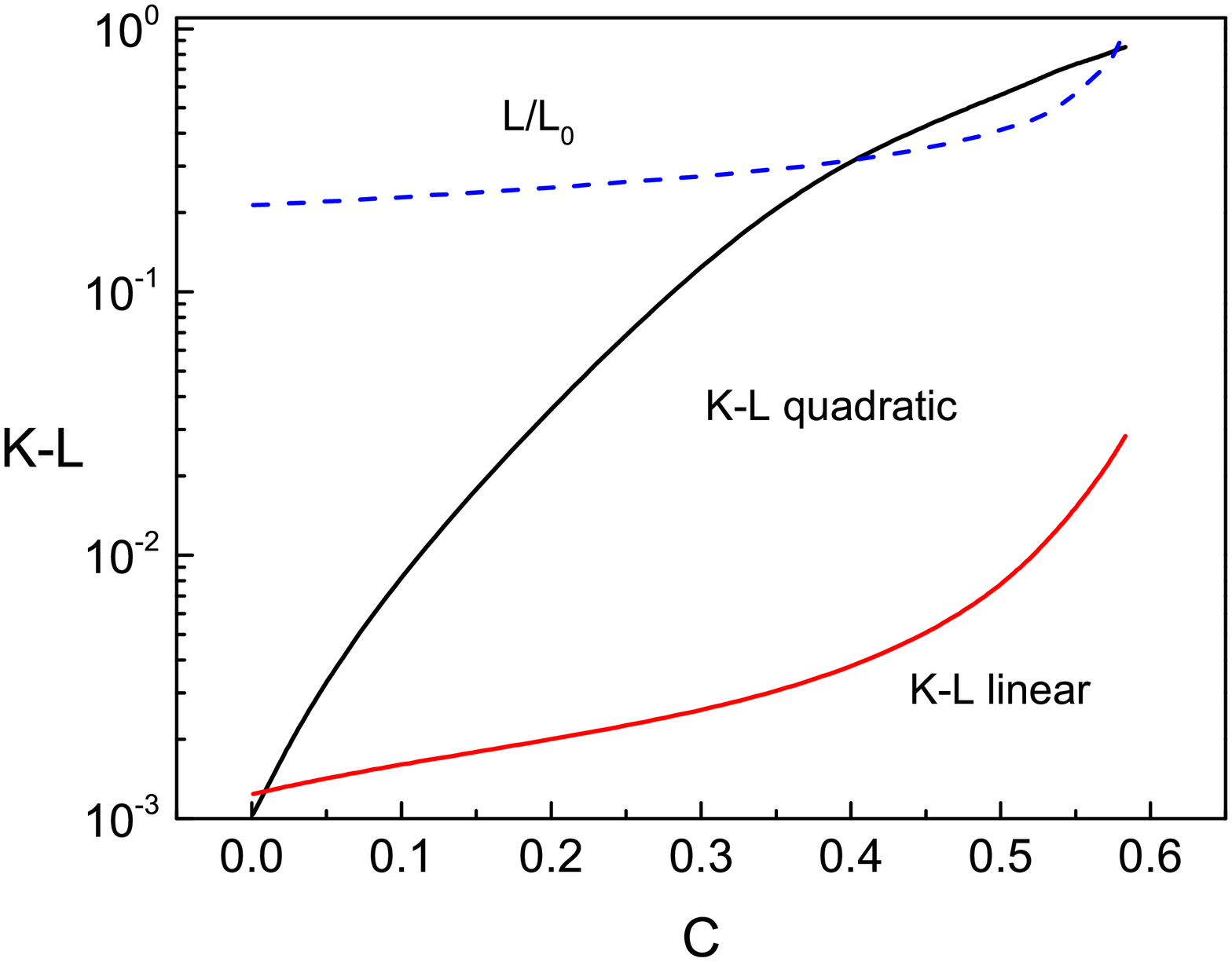}\label{dkl2:f2}}
  \caption{Normalized K-L distance for different values of $p_d$ on SWNs and with linear and quadratic 
dynamics. a) as a function of $L/L_o$, with $L_o$ the mean path length of the ordered network. The 
dashed line is the value of the clustering $C$; b) as a function of $C$. The dashed line is the value of $L/L_o$.}
\end{figure*}

\subsubsection{SWN}

While the construction of SW networks is similar to its regular version and also presents 
an analogous behavior in terms of the values of $C$ and $L$ as a function of the disorder 
coefficient, there exists an apparent difference. In addition to the effects of the topology on the 
formation of consensus due to the change of $C$ and $L$ we need to consider the effects 
that could arise due to the change in the degree distribution.
As in the previous case, we observe a transition towards consensus that can be much better 
quantified by studying the behavior of the K-L distance.
The plots in Figs. \ref{dkl2:f1} and \ref{dkl2:f2} show the dependence of the K-L distance on $C$ 
and $L$. In this case, we observe that the growth of the K-L distance behaves differently than 
the analogous situation in $k$-SWN for the quadratic dynamics, the one that promotes consensus 
formation. This difference must be due to the combined effect of an increase in $L$ and $C$ and a 
change in the degree distribution.

\begin{figure}[ht]
\begin{center}
\includegraphics[width=\columnwidth]{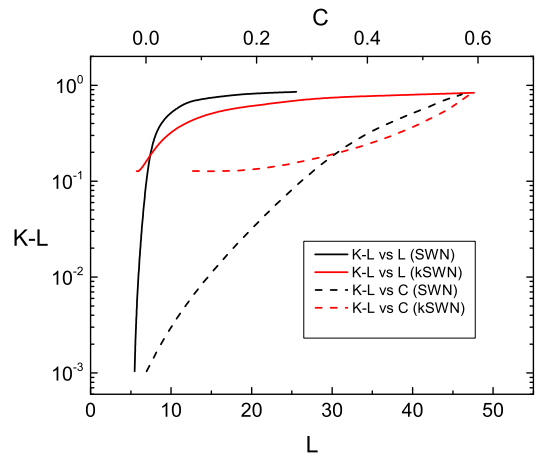}
\end{center}
\caption{Comparison of the normalized K-L distance as a function of $L$ (lower $x$-axis,) and $C$ (upper 
$x$-axis).}
\label{klcont}
\end{figure}

Indeed, if we compare both families of networks in a unique plot it is more evident 
that the spreading of the degree distribution goes against the formation of consensus. We show 
this in Fig \ref{klcont}, where we contrast the values of the K-L distance for the same values of 
clustering $C$ and mean path length $L$ for both types of networks. The figure shows that for the 
same values of clustering   or path length, SWN networks start to fail in reaching consensus  once 
the degree distribution has adopted a clear binomial profile, corresponding to lowest values of $C$ 
and $L$.
A natural next step is to consider networks with more spanned degree distribution. A candidate for 
that are the Scale Free networks. 

\begin{figure}[ht]
\begin{center}
\includegraphics[width=\columnwidth]{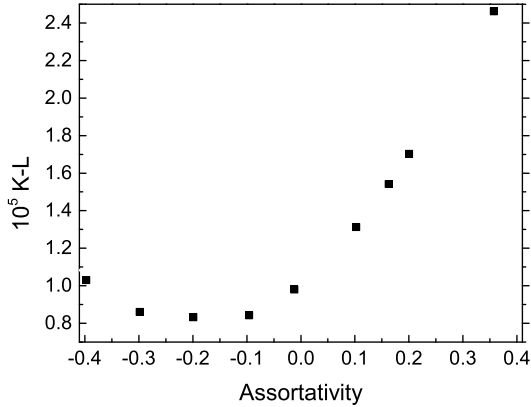}
\end{center}
\caption{Normalized K-L distance for different values of assortativity $a$ on SF networks with quadratic 
dynamics.}
\label{sf1}
\end{figure}

\begin{figure}[ht]
\begin{center}
\includegraphics[width=\columnwidth]{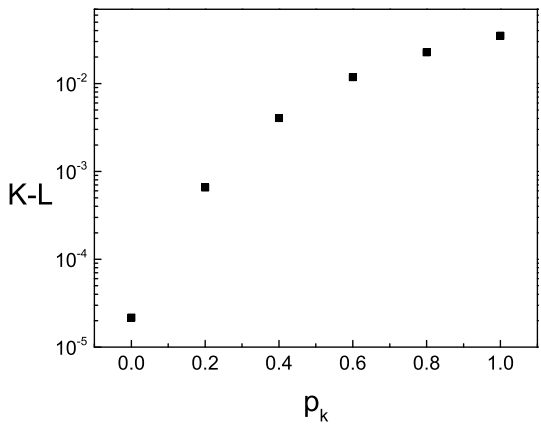}
\end{center}
\caption{Normalized K-L distance as a function of $p_k$ for a SF network with $a=0.4$, quadratic dynamics and 
degree dependent navigation strategy.}
\label{navi}
\end{figure}

\subsubsection{SF Networks}
So far we have studied SW networks with different degrees of disorder. The difference between SWN 
and $k$-SWN evidenced the effect of the heterogeneity of the degree distribution on the dynamics, 
that goes against the formation of consensus.  SF networks owe their name to their particular 
degree distribution. We choose to analyze the behavior of the 
consensus emergence  on these networks and compare it with our previous results. But SF 
networks are not only characterized by a degree distribution with a tail that falls with a power 
law. 
Among other properties,  two networks might have the same degree distribution but present different assortativities.
We choose to study the effect of this quantity in order to unveil the role of hubs or highly 
connected clusters of nodes in the obtained results. 
The calculated values of  the K-L distance are low, as expected from previous results with 
highly disordered SW networks. The interesting feature of this case  is associated to the behavior 
of the K-L distance  as a function of the assortativity. We find that the emergence of a 
weak consensus corresponds to highly assortative or disassortative networks while neutral 
networks, that can also be linked to certain degree of heterogeneous structure,  lead to its 
complete inhibition. This can be observed in Fig. \ref{sf1}. In any case we obtain much smaller values than
in the small world network, in agreement with the result that long tails in the degree distribution 
impede the formation of consensus.

For this family of networks we have also studied a different navigation dynamics.
Instead of choosing the next step at random, the choice pointed to the neighbor with the highest 
degree
with probability $p_k$, and with complementary probability to any node in the neighborhood.  
The case $p_k=0$ corresponds to the previous one. While for neutral and disassortative SF networks 
we did not find any apparent difference, we 
noticed 
that this last strategy has an interest effect on assortative SF networks. As shown in Fig. 
\ref{navi}, the consensus level increases with $p_k$, as the navigation strategy concentrates many 
of the walkers in a bounded subnet corresponding to those nodes with the highest degrees, allowing 
for the occurrence of much more interactions.

\subsection{Transient dynamics}

Most of the features observed once the totality of agents have arrived to the target can be 
explained by analyzing the transient dynamics. We will discuss some of the most relevant 
observations in the following sections, starting with those aspects that only characterize the way 
the walkers navigate the network to finally address a characterization of the interactions rate, 
close linked to the consensus emergence.

\begin{figure*}[!ht]
  \centering
  \subfloat[]{\includegraphics[width=0.5\textwidth]{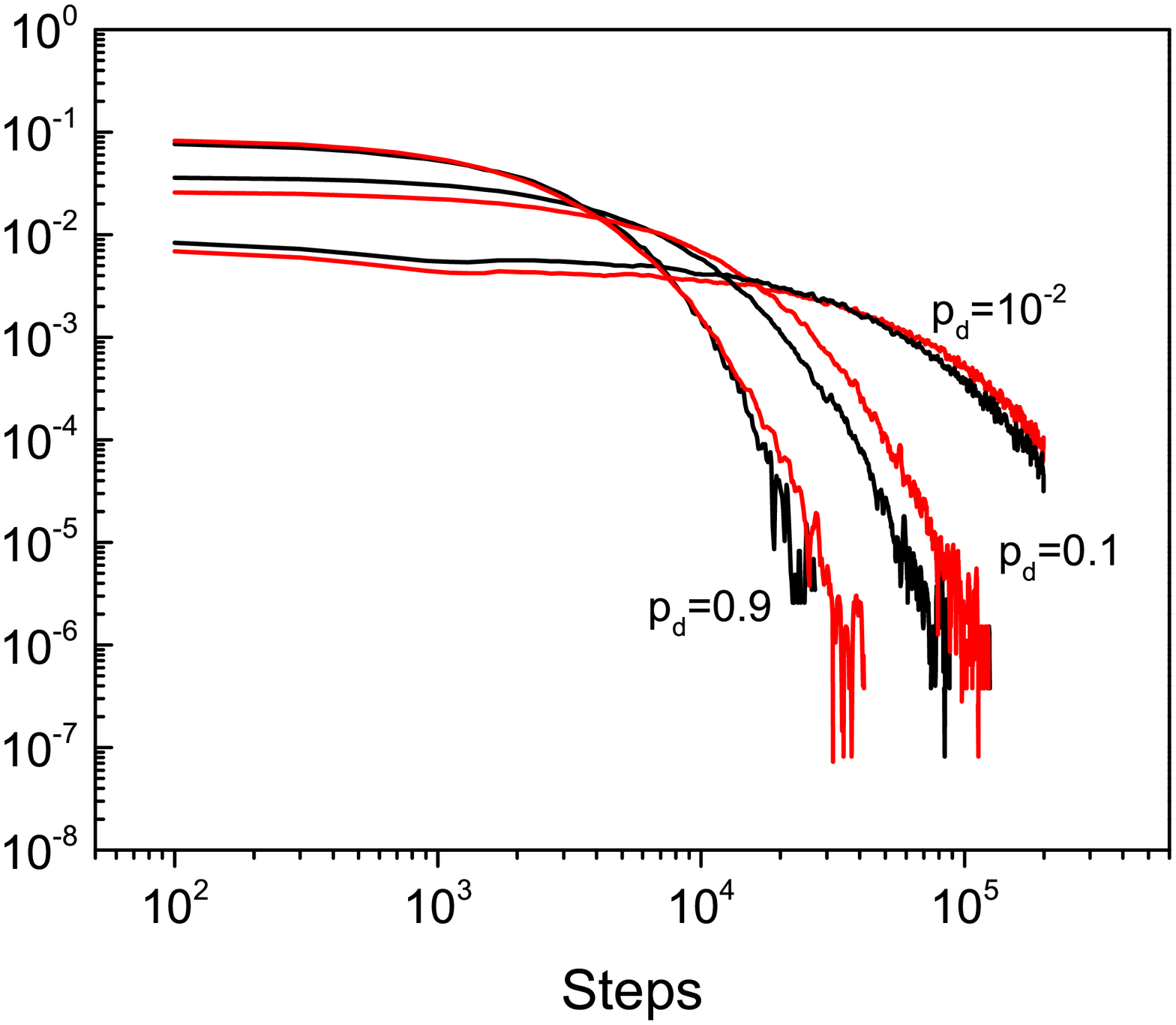}\label{steps:f1}}
  \hfill
  \subfloat[]{\includegraphics[width=0.5\textwidth]{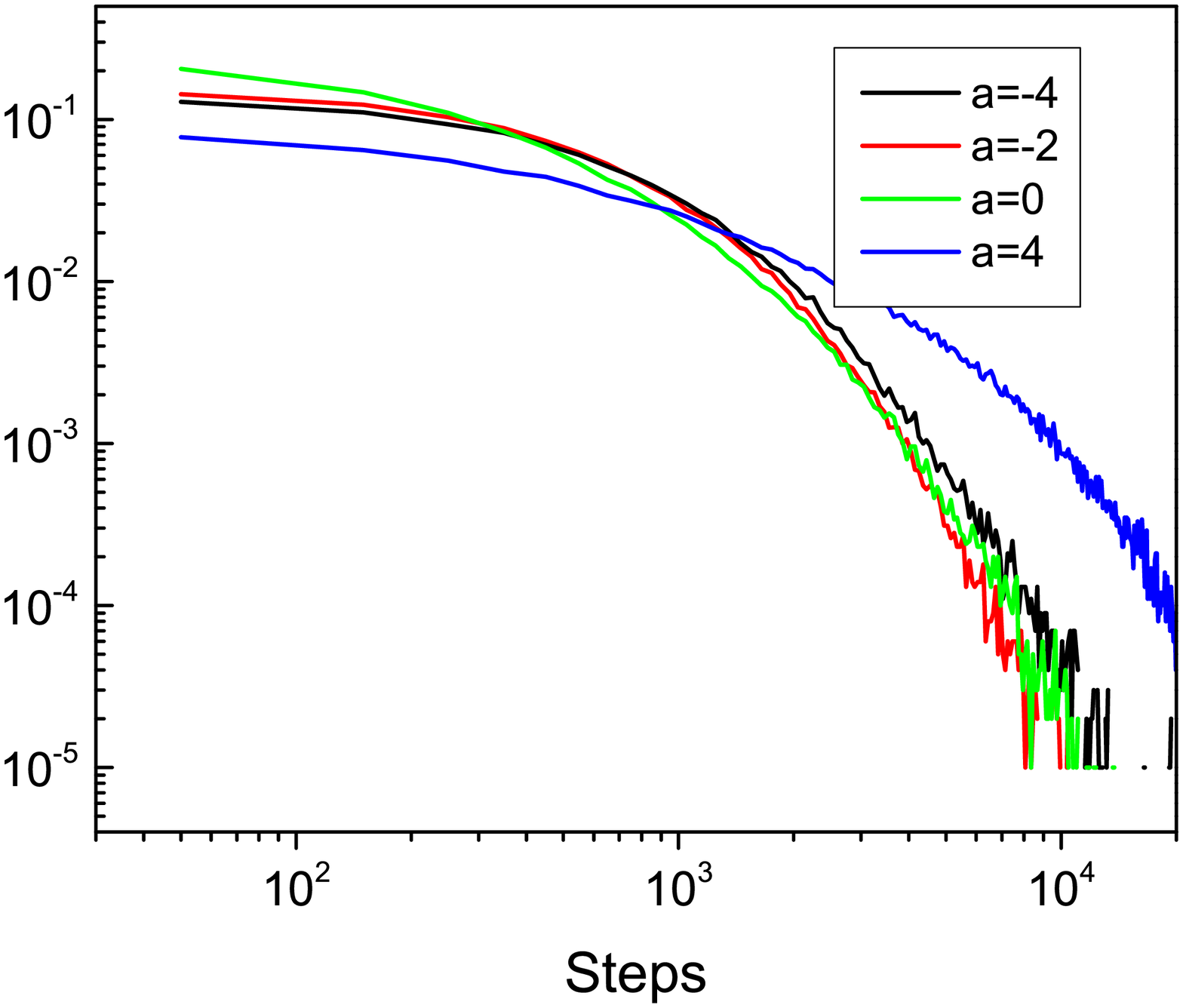}\label{steps:f2}}
  \caption{Step distribution for a) $k$-SWN (black) and 
SWN (red) , and b) SF networks}
\end{figure*}

\subsubsection{Distribution of walking length}

The process of consensus formation requires that the agents change their opinion in their way to
their target destination. By studying the amount of steps a given agent needs to make from the 
origin to the destiny and knowing the  probability of change of opinion at 
each step  we can estimate the probability that an opinion has been altered.
Though at first glance the results show us a trivial  correlation between the length of the walks 
and the emergence of consensus or collapse of opinions, a closer analysis reveals same features 
that need to be observed with care.
Fig. \ref{steps:f1} shows the distribution of the number of steps (or walking length) for  $k$-SWN 
and SWN  for different disorder degrees. In  Fig. \ref{steps:f2} we plot the same for SF networks. 
As can be observed, there is a dramatic reduction in the number of required steps as the network 
gets more disordered.  Despite the mentioned association between 
consensus emergence and longer walks, this plots shed no light about why the results for highly
disordered $k$-SWN and SWN networks are so different, which have step distributions that are almost 
identical.  

These last curves are very similar to those observed for SF networks.
Besides, the plot corresponding to SF networks shows a non monotonic dependence on the 
assortativity. These results are consistent with what have been already shown in the previous 
section. Consensus formation is harder  when the underlying network is only barely disassortative.

It is important to note that these results are independent of the choice of the dynamics and 
reveal only the topology of the walks. Though  the distribution of walking length is directly 
linked to the performance of the networks in promoting consensus, it is not the 
only relevant feature, as has been already anticipated and will be further discussed in the 
following sections.

\subsubsection{Buffer dynamics}

\begin{figure*}[!ht]
  \centering
  \subfloat[]{\includegraphics[width=0.5\textwidth]{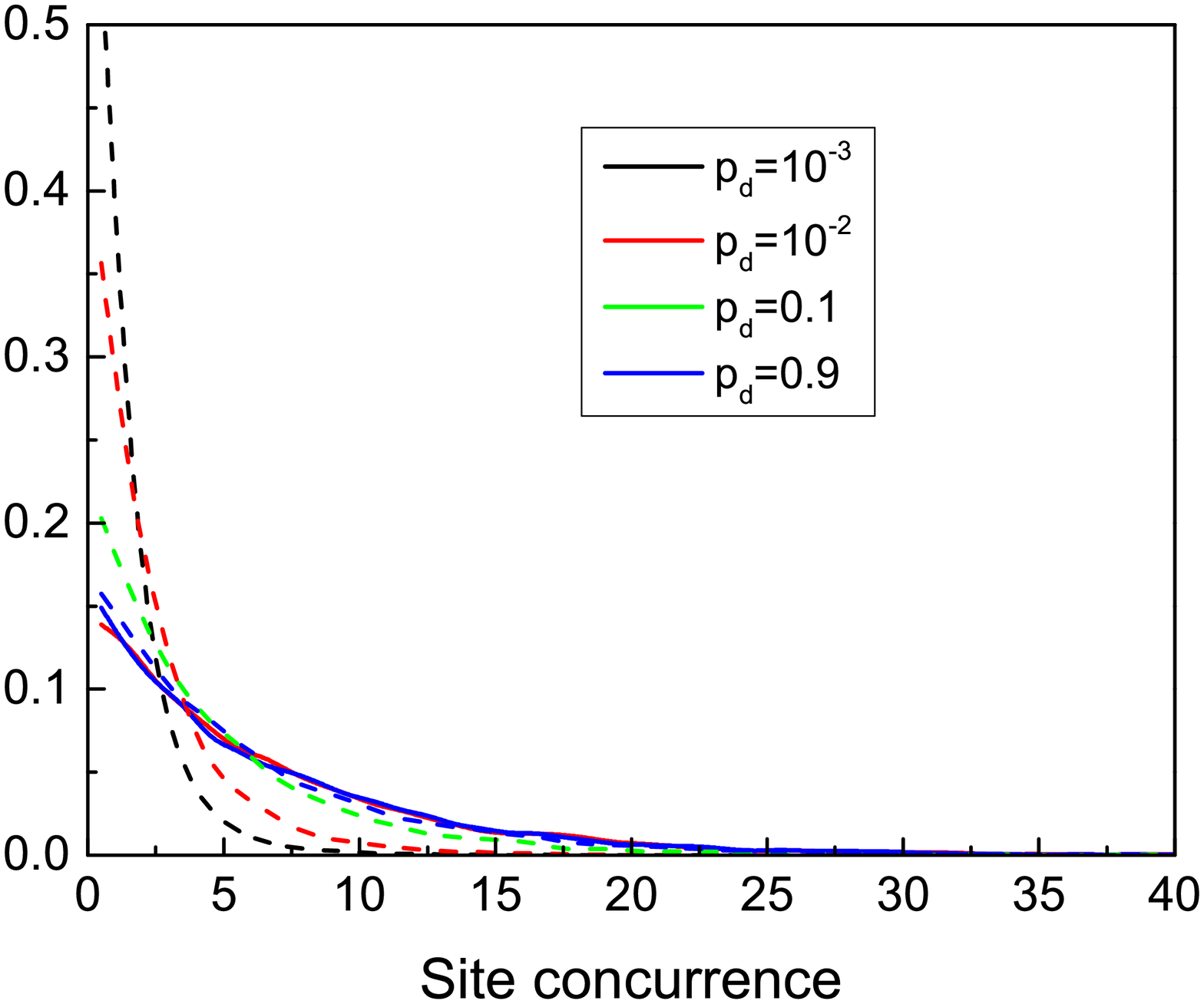}\label{buf:f1}}
  \hfill
  \subfloat[]{\includegraphics[width=0.5\textwidth]{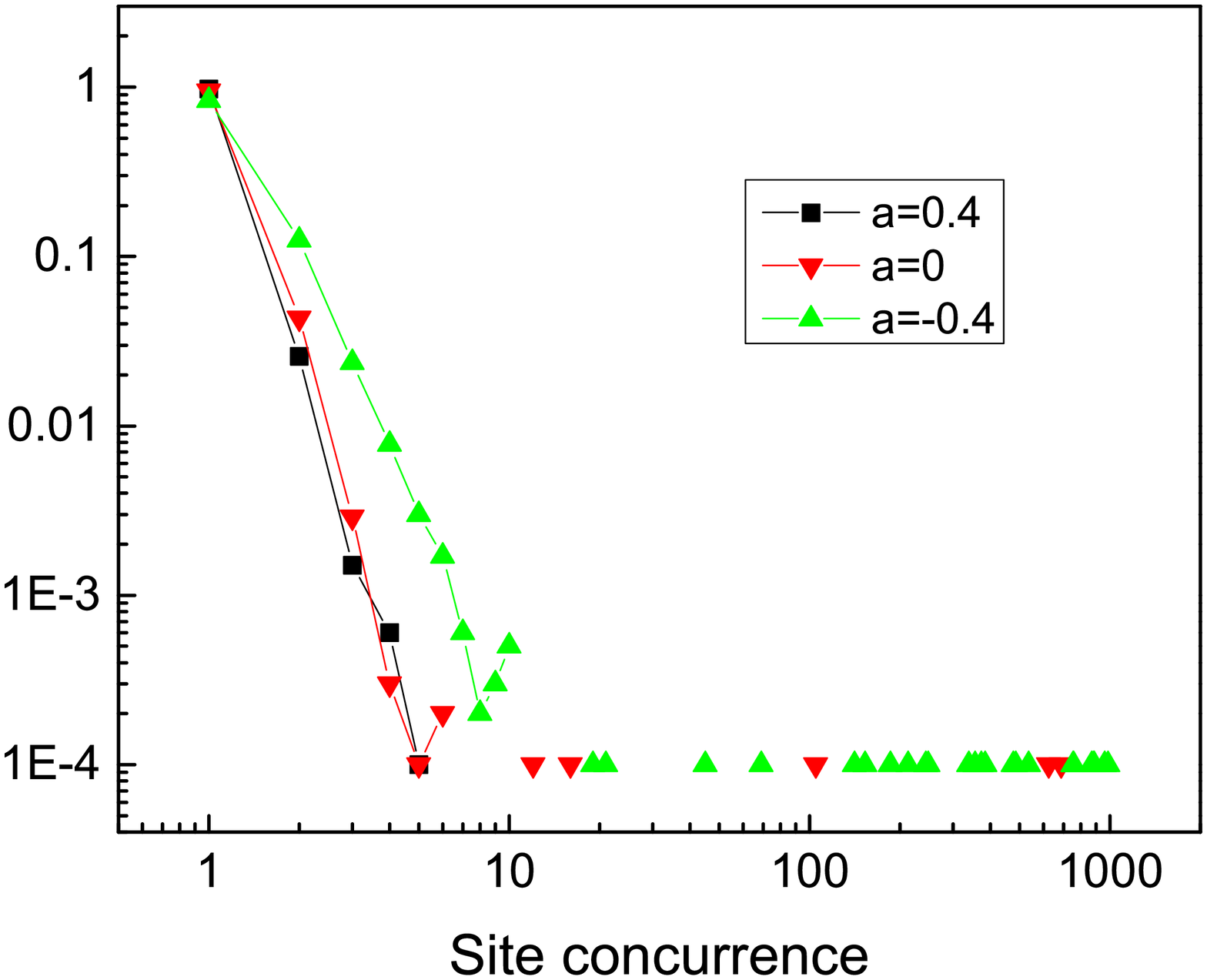}\label{buf:f2}}
  \caption{Distribution of the number of agents per node when 40\% of packages has been delivered 
for  a) $k$-SWN (solid) and SWN (dashed) , and b) SF networks}
\end{figure*}

From the moment the agents start to reach their destination, the initial  distribution of the number 
of agents in the nodes changes its shape, starting from a Gaussian-like distribution to turn into one  
with a high number of nodes with a low number of agents and a few with a high number of them. The 
rate of 
change of this distribution depends on the topology of the network. However, if we analyze 
these distributions as a function of the amount of arrived agents instead of the spent time we 
find, in the case of  $k$-SWN  that they are almost the same, as shown in Fig \ref{buf:f1} where we 
plot the distribution of agents per node once 40\% of the walkers have reached destination for 
several values pf $p_d$. Also shown in the same figure is the case for SWN. This time, the 
distributions do not overlap. 
We can observe that the curve corresponding to  all the $k$-SWn and to 
the SWN with $p_d=10{-3}$ are overlapped. On the contrary, SWN with different disorder degrees 
present different behaviors. The last is also true for for SF networks with different 
assortativity values, as shown in Fig \ref{buf:f2}.

We start to sketch here a first explanation to the different behavior of the consensus on $k$-SWN 
and  SWN. Together with a non linear interaction dynamics, the consensus requires the encounter 
and interaction of walkers to occur. The plots show us that the crowded sites are more abundant in 
$k$-SWN  than in the others networks. This last fact promotes consensus formation.

\subsubsection{Opinion change dynamics}

\begin{figure*}[!ht]
  \centering
  \subfloat[]{\includegraphics[width=0.5\textwidth]{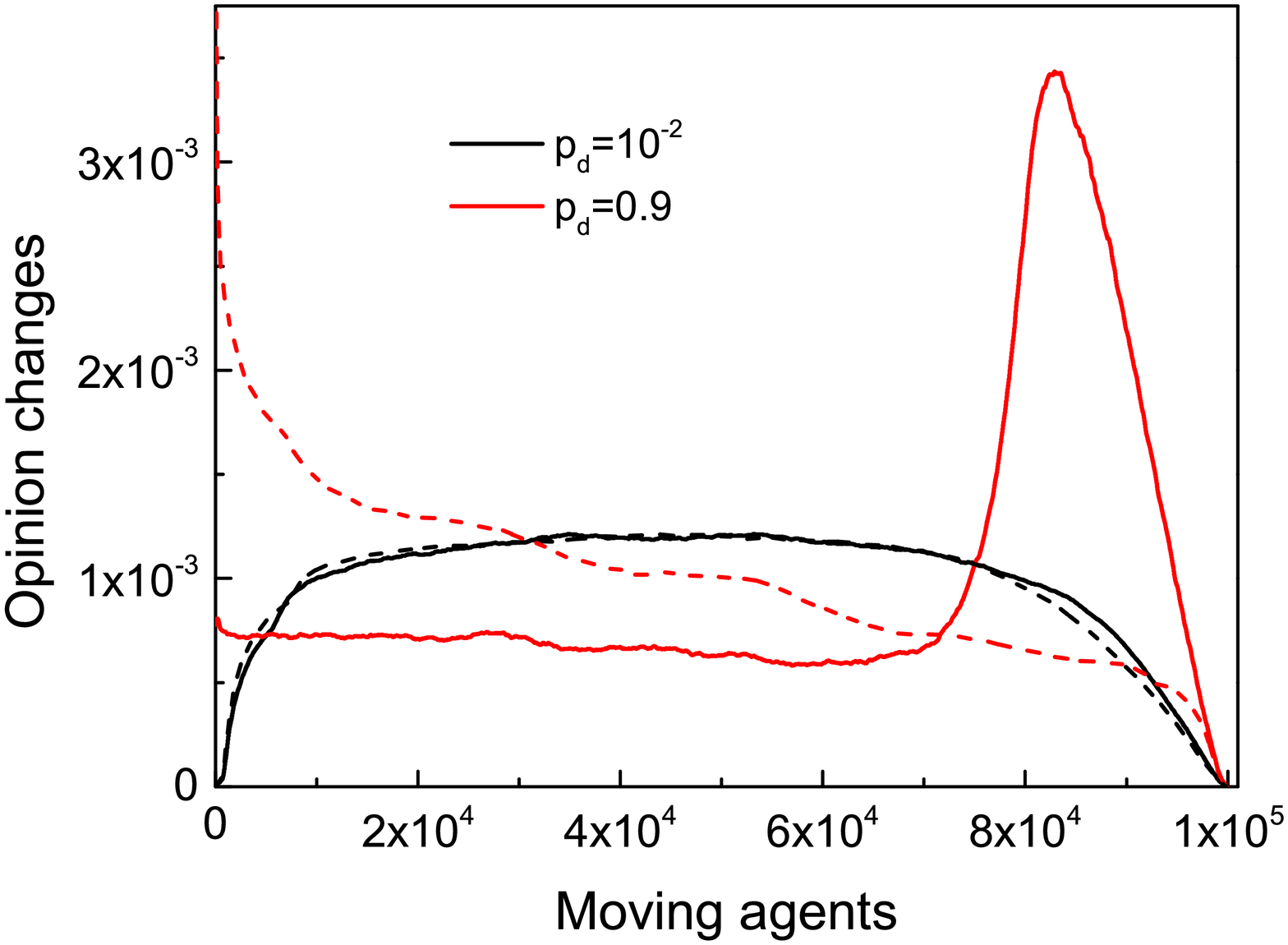}\label{errd:f1}}
  \hfill
  \subfloat[]{\includegraphics[width=0.5\textwidth]{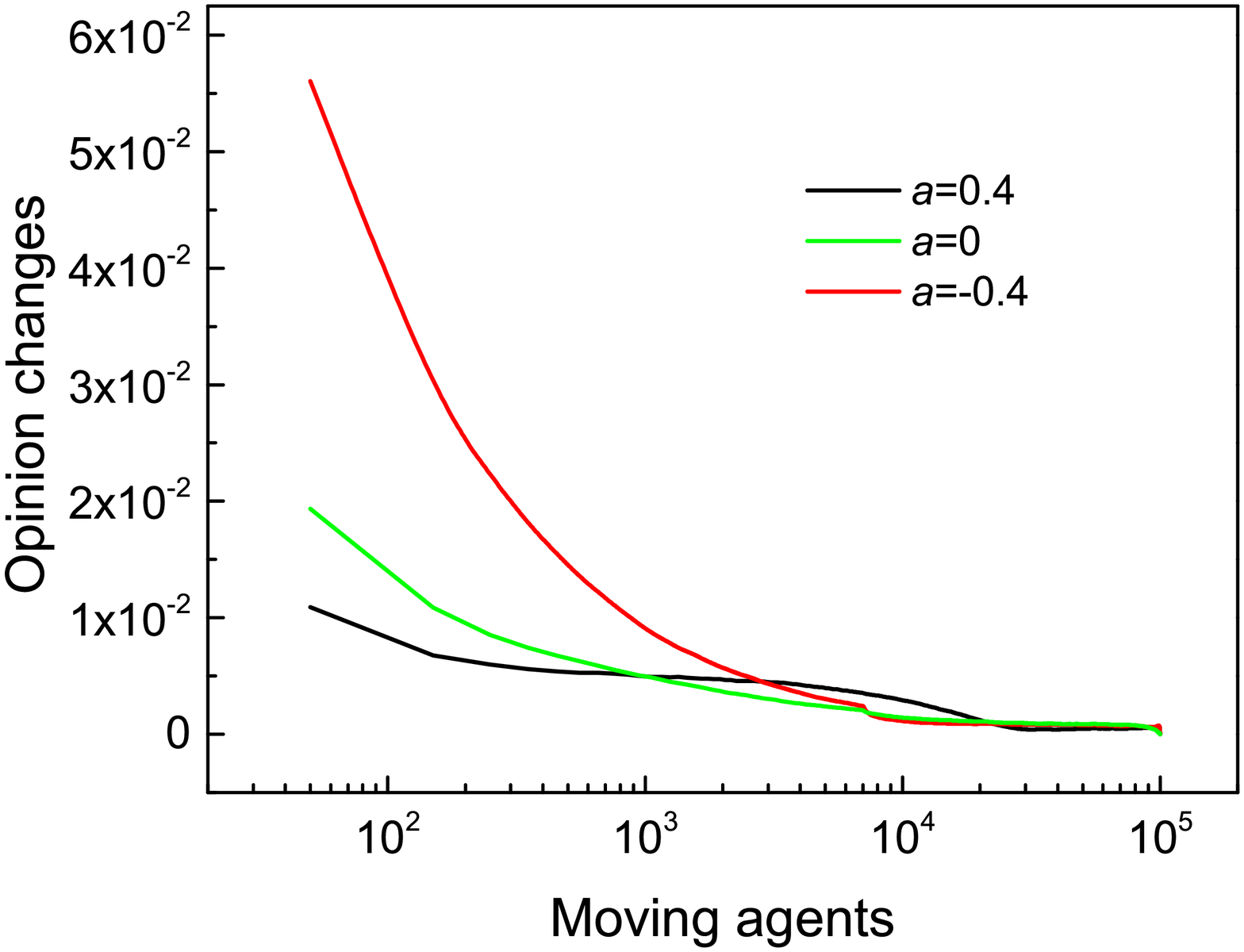}\label{errd:f2}}
  \caption{Changes of opinion as a function of the number of moving agents for  a) $k$-SWN (solid) and 
SWN (dashed) , and b) SF networks}
\end{figure*}

A natural question is at what stage of the dynamics occur the most part of the change of opinions.
For that, we have recorded the rate of change occurrence and plotted it against the amount of 
agents that reached destination. While $k$-SWN and SWN show the same dynamics for low values of $p_d$, 
they are very different for higher disorder degrees. This is shown in Fig \ref{errd:f1}.
In highly disordered $K$-SWN the changes occur at the beginning of the dynamics (high number of 
mobile agents), while for SWN it is just the opposite. This is
counterintuitive because the natural guess is to expect that most of the interactions occur when a 
high number 
of agents are still walking, as with $k$-SWN . In turn, SF networks behave as highly 
disordered SWN, but with a change rate much more concentrated at low values of walking agents, as 
shown in Fig \ref{errd:f2}. 

\section{Discussion}

In the present work we have considered a model for the emergence of consensus in which the involved 
individuals are not influenced by static neighborhoods but participates from several dynamical fora 
where several individual gather and discuss. Different families of networks serve to mimic a 
variety of underlying topologies for these dynamical neighborhoods. The aim of of such election is 
to analyze how the particular properties of these topologies affect the emergence of consensus. By 
consensus we understand the collapse of different opinions into a single one. In turn, the opinions 
of the individuals is modeled as a scalar, ranging from 1 to 10 and without any notion of 
closeness between different opinions.
That means that the opinion with value equal to 1 is not closer to the opinion with value equal to 2 than it is to 
another with value equal to 10.
Our results suggest that an heterogeneous topology inhibits the formation of consensus. The 
heterogeneity in the considered networks is associated to a disordered structure, an extended  
degree distribution and the lack of correlation between nodes degree. 

We have shown that a linear interaction dynamics, that consist only in a replacement of and idea by 
other one does not promote consensus. On the contrary, what we called quadratic  dynamics 
favors the propagation of coincidences, leading to consensus. 
We have also considered some alternative dynamical rules with results that are analogous to what we 
have obtained for the quadratic one. Though they were not discussed in this work they are worth 
mentioning. These dynamics are  
\begin{itemize}
\item{Higher order $n$:} Analogous to the quadratic taking $n$ opinions 
among the present in the node. 
\item{Poissonian:} The number $n$ considered in the previous case is taken from a Poisson 
distribution.
\item{Uniform:} As in the previous case, with a $n$ uniformly distributed.
\end{itemize}

As mentioned before, the results are qualitatively the same to which we obtained with the 
quadratic dynamics, suggesting that the quadratic case represents a generic non-linear interaction. 
However it must me notice that for a purely case  that finite size effects become more severe  than
for the quadratic, because the change of opinion now requires that there are at least three agents in 
node.

The present results can be also interpreted in a different framework such as  errors in the 
transmission of information. When considering information transfer, the occurrence of errors during 
the process is of fundamental interest. Therefore, any realistic process of transfer information 
should include the possibilities of making mistakes that can degrade the performance of the system. 
These mistakes can be assumed to take the form of random processes that occur in parallel to the 
transmission of information. It is usually assumed that the mistakes affect the directions to which 
the packets of information are being delivered. For instance, a package could be delivered to a 
random networks node instead of its desired destination or the topology of the network could be 
perturbed for all the messages that go through a given region. These aspects have been analyzed  in 
ref. \cite{czaplicka}. Interestingly, 
it is found that in some cases this kind of noise can enhance transfer of information via a phenomenon similar 
to stochastic resonance, with a non trivial interaction between the two types of noise. 
But we can have also the  possibility that noise does not affect the destination of the messages but its content.
This is the equivalent of what here, with the opinions of the agents take the place of the content 
of the messages.


\begin{thebibliography}{}

\bibitem{bocca} S. Boccaletti , V. Latora, Y. Moreno, M. Chavez, D.-U. Hwang,
Phys. Rep. \textbf{424}, 175–308 (2006)



\bibitem{kup1} M. Kuperman, G. Abramson,
Phys. Rev. Lett.  \textbf{86}, 2909.  (2001)

\bibitem{kup2} G. Abramson and  M. Kuperman
Phys.  Rev.  E \textbf{63}, 030901R (2001)

\bibitem{kup0} M.N. Kuperman
Phys. Rev. E \textbf{73}, 046139  (2006)

\bibitem{jack} M.O. Jackson, B. Rogers, Advan. Theoret. Econ. \textbf{7}, 1–13  (2007)

\bibitem{lopi} D. L\'opez-Pintado, 
Games and Economic Behavior \textbf{62},  573–590 (2008)

\bibitem{garde} J. G\'omez-Garde\~nes, V. Latora
Phys. Rev. E \textbf{78}, 065102(R) (2008)

\bibitem{wang-shang2015} H. Wang, Shang, L. 
Physica A \textbf{421}, 180-186 (2015)

\bibitem{deGroot1974}  M. H. DeGroot, 
Journal of the American Statistical Association \textbf{69}, 118-121 (1974)

\bibitem{Berger1981} R. L. Berger, 
Journal of the American Statistical Association \textbf{76}, 415-418 (1981)

\bibitem{GilardoniClayton1993}  G. L. Gilardoni,  M. K. Clayton, 
The Annals of Statistics, 391-401 (1993) 

\bibitem{holley1975} R. A. Holley,  T. M.  Liggett,
The annals of probability, 643-663 (1975)

\bibitem{sznajd-weron2000} K. Sznajd-Weron, J. Sznajd, 
International Journal of Modern Physics C \textbf{11}, 1157-1165  (2000)

\bibitem{demarzo2003} P. M. DeMarzo, D.  Vayanos, J. Zwiebel,
The Quarterly Journal of Economics \textbf{118}, 909-968  (2003)

\bibitem{fazeli2011}  A. Fazeli, A. Jadbabaie, 
In Decision and Control and European Control Conference (CDC-ECC), 2011 50th IEEE Conference on 
(pp. 2341-2346). (IEEE, 2011).

\bibitem{sato} M. Barthélemy, A. Barrat, R. Pastor-Satorras,  A. Vespignani
Phys. Rev. Lett. \textbf{92}, 178701 (2004).

\bibitem{huan} L. Huang, K. Park,  Y. C. Lai, Phys. Rev. E \textbf{73}, 035103-R (2006)







\bibitem{watt} D.J. Watts, S.H. Strogatz, Nature \textbf{393}, 440 (1998)

\bibitem{kup3} M. N. Kuperman, S. Risau-Gusman, Eur. Phys. J. B \textbf{62}, 233–238 (2008)

\bibitem{barra} A. Barrat, M Weigt, Eur. Phys. J. B. \textbf{13}, 547–560 (2000)

\bibitem{bara} A. L. Barab\'asi, R. Albert,  Science \textbf{286}, 509 (1999)

\bibitem{xulv} R. Xulvi-Brunet,  I. M. Sokolov, 
Phys. Rev. E \textbf{70}, 066102 (2004)

\bibitem{newa} M.E.J. Newman.  Phys. Rev. Lett. \textbf{89}, 208701 (2002) 


\bibitem{kim} B.J. Kim, C.N. Yoon, S.K. Han, H. Jeong, Phys. Rev. E \textbf{65},  27103 (2002)

\bibitem{noh} J.D. Noh, H. Rieger, Phys. Rev. Lett. \textbf{92}, 118701  (2004)

\bibitem{adam} L. Adamic, R.M. Lukose, A.R. Puniyani, B.A. Huberman, Phys. Rev. E \textbf{64},
46135  (2001)



\bibitem{kul} S. Kullback, R. A. Leibler,  Annals of 
Mathematical Statistics \textbf{22}, 79–86 (1951)


\bibitem{czaplicka} A. Czaplicka, J. A. Holyst , P. M. A. Sloot,
Scientific Reports \textbf{3} 1223 (2013)


\end{thebibliography}
\end{document}